\documentclass[twocolumn,abs,prb,showpacs]{revtex4-1}

\usepackage{graphicx}
\usepackage{dcolumn}
\usepackage{float}
\usepackage{amsmath}
\usepackage{bm}
\usepackage[mathscr]{euscript}

\newcommand{\comment}[1]{}

\begin{document}


\title{Particle-Hole Asymmetry in Gapped Topological Insulator Surface States}

\author{C. J. Tabert$^{1,2}$}
\author{J. P. Carbotte$^{3,4}$}
\affiliation{$^1$Department of Physics, University of Guelph,
Guelph, Ontario N1G 2W1 Canada} 
\affiliation{$^2$Guelph-Waterloo Physics Institute, University of Guelph, Guelph, Ontario N1G 2W1 Canada}
\affiliation{$^3$Department of Physics, McMaster University,
Hamilton, Ontario L8S 4M1 Canada} 
\affiliation{$^4$Canadian Institute for Advanced Research, Toronto, Ontario M5G 1Z8 Canada}
\date{\today}

\begin{abstract}
{We consider the combined effect of a gap and a Zeeman interaction on the helical Dirac fermions which exist on the surface of a topological insulator.  Magneto-optical properties, the magnetization, Hall effect and the density of states are considered with emphasis on the particle-hole asymmetry which arises when a subdominant Schr{\"o}dinger piece is included along with the dominant Dirac part of the Hamiltonian.  When appropriate, we compare our results with those of a single valley gapped graphene system for which Zeeman splitting behaves differently.  We provide a derivation of the phase offset in the magnetic oscillations brought about by the combined effect of the gap and Schr{\"o}dinger term without requiring the semiclassical Onsager quantization condition.  Our results agree with previous discussions based on semiclassical arguments.
}
\end{abstract}


\maketitle

\section{Introduction}

A topological insulator (TI)\cite{Hasan:2010, Qi:2011} is a material which is insulating in the bulk but posses a metallic spectrum of surface states.  These are expected to have high mobility as they are protected by the topology of the band structure.  The surface states of Bi$_2$Se$_3$ have a single Dirac cone\cite{Hsieh:2009} centred at the $\Gamma$-point of the two-dimensional (2D) surface Brillouin zone.  Other materials can have a different odd number of such points.  As an example, samarium hexaboride has three\cite{Roy:2014}.  Such a new phase of matter comes about through a band inversion caused by a strong spin-orbit coupling.  At present, these systems are extensively studied due to their novel physics and their possible applications in devices such as quantum computing\cite{Fu:2008} and photonics\cite{Plucinski:2002}.  An important observation is the phenomenon of spin-momentum locking\cite{Hsieh:2009} established by spin- and angular resolved photoemission spectroscopy (ARPES).  This shows that the in-plane component of the spin is perpendicular to the in-plane momentum.  A gap can also be introduced at the Dirac point through doping of the TI surface with magnetic (time-reversal-breaking) impurities.  In the work of Chen et al.\cite{Chen:2010}, a gap of $\Delta\sim 7$ meV was observed through ARPES measurements on a sample of (Bi$_{0.99}$Mn$_{0.01}$)$_2$Se$_3$ with $1\%$ Mn substitution on the Bi site.  Another avenue to create a gap is to make the TI ultra thin with a distance between the top and bottom surface of the order of the extent in space (perpendicular to their plane) of the surface states\cite{Lu:2010,Linder:2009,Shan:2010}.  Tunnelling between the top and bottom surfaces gaps the electronic spectrum.  Of course, in this case, both surfaces can contribute to a particular phenomenon and a sum over both cones (which have gaps of opposite sign) is needed as has been discussed recently by  Yoshimi et al.\cite{Yoshimi:2015} and Zhang et al.\cite{Zhang:2015} in the context of the quantum Hall effect.  Neupane et al.\cite{Neupane:2014} find that varying the quantum-tunnelling gap in ultra-thin films leads to a modulation of the in-plane spin texture.  While the spin-momentum locking remains, the in-plane component of spin can itself be decreased with decreasing thickness.  Other such studies include work by Tahir et al\cite{Tahir:2013b} on the oscillations expected in the quantum capacitance of thin films.  For simplicity, we will consider a single gapped Dirac cone; however, the results applicable to thin films can be obtained by assembling two such cones with the appropriate relationship between their respective gaps.

Particle-hole asymmetry is an important feature of the surface-fermion band structure of a TI.  This arises from a subdominant quadratic-in-momentum Schr{\"o}dinger term with mass $m$ in addition to the dominant linear-in-momentum Dirac term with Fermi velocity $v_F$.  The relativistic piece has its origin in the strong spin-orbit coupling and involves the Pauli spin-matrices $\hat{sigma}_x$ and $\hat{\sigma}_y$.  This is distinct from the case of graphene where the matrices exist in pseudospin space and the energy levels are degenerate in spin\cite{Gusynin:2006b}.  With the application of a magnetic field perpendicular to the plane, the Landau levels (LLs) are split by the Zeeman field and form two sublevels with a definite $s_z$ component of spin $\pm\hbar/2$\cite{Gusynin:2006b}.  This contrasts with the case of the TI for which only the $N=0$ level is fully spin polarized in $s_z$\cite{Vazifeh:2012} while the average $s_z$ for all the other levels is small in comparison.  For the presently studied TI, the Dirac magnetic energy scale at one Tesla is an order of magnitude larger than the Schr{\"o}dinger scale.  However, at $B=36$T (for example) the two scales differ by less than a factor of two.  Even when $B$ is small, the Schr{\"o}dinger term can lead to novel effects such as the splitting of the optical absorption lines seen in the real part of the AC conductivity\cite{ZLi:2013}. Thus, this term must be included in any complete theoretical treatment and is important when considering experimental data.  In particular, optical data gives valuable information on the dynamics of the surface charge carriers in a TI as demonstrated in the experimental work by LaForge et al.\cite{LaForge:2010} and Orlita et al.\cite{Orlita:2015} among others.

Our paper is organized as follows: in Sec.~II, we specify our low-energy Hamiltonian involving Schr{\"o}dinger-Dirac kinetic energy and the gap term $\Delta\hat{\sigma}_z$ which does not commute with the rest of the Hamiltonian.  The eigenenergies and eigenfunctions which arise in the presence of a magnetic field are worked out; from these, the average value of $s_z$ for each level is determined.  We also consider a related system written in pseudospin space.  In this case, the wave-functions are simultaneous eigenstates of the Hamiltonian and $\hat{s}_z$.  Thus, each LL is fully spin polarized in the $z$ direction.  In Sec.~III, we begin our discussion by considering the integrated density of states up to energy $\omega_{\rm max}$ (as measured from the Dirac point).  We emphasize the effect of particle-hole asymmetry in a TI and contrast this with the case of a single valley of a graphene-like system which shows no asymmetry in the absence of a gap.  We also discuss the modifications brought about by a finite $\Delta$ and by Zeeman splitting.  Section~IV contains formulas for the magneto-optical conductivity.  Both the longitudinal and the transverse Hall conductivities are considered.  Again, particle-hole asymmetry is emphasized.  We find a comparison with the single-valley graphene-like system to be helpful.  Circular dichroism is discussed.  In Sec.~V, we examine the magnetization of the metallic surface states.  Through the Streda relation\cite{Smrcka:1977,Wang:2010}, we relate its derivative with respect to chemical potential ($\mu$) to the quantization of the Hall plateaus.  As expected, we find complete agreement between this and our conductivity formula.  We find that the quantization remains half-integral (i.e. $\pm 1/2, \pm 3/2, \pm 5/2,...$) which is characteristic of relativistic fermions (with due consideration for the degeneracy factors).  The value of $\mu$ at which a new step occurs is changed with the value of Schr{\"o}dinger mass, $\Delta$ and Zeeman splitting.  In contrast to graphene, Zeeman interactions do not introduce additional steps\cite{Gusynin:2006b} between those present when the effect is neglected.  It is important to realize that when the Dirac term in our Hamiltonian is dominant, this describes a TI and the Hall quantization is relativistic.  When the Schr{\"o}dinger piece dominates, the Hall quantization is non-relativistic and the Hamiltonian applies to the usual spintronic materials.  In Sec.~VI, we consider the magnetic oscillations and in particular, emphasize the phase offset in the usual semiclassical expression.  This quantity is 1/2 for non-relativistic Schr{\"o}dinger particles and 0 for relativistic Dirac fermions.  In the presence of a finite mass and magnetic gap, we find that the offset is changed to $\gamma=-\Delta(1+g)/(2mv_F^2)$, where $g$ is the Zeeman strength.  This reduces to the relativistic result for either $\Delta=0$ or $m\rightarrow\infty$.  The phase offset depends on the Zeeman term through the coupling coefficient $g$.  Except for this factor, our quantum mechanical results agree with a previous result\cite{Wright:2013} which relied on semiclassical arguments\cite{Luk:2004,Fuchs:2010,Taskin:2011,Mikitik:2012,Ando:2013,Raoux:2014,Kishigi:2014}.  Fuchs et al.\cite{Fuchs:2010} start with the Bohr-Sommerfeld quantization condition suggested by Onsager\cite{Onsager:1952} and include a correction in the band energies due to the magnetization $\mathcal{\bm{M}}(\bm{k})$ induced by the external magnetic field $B$ of the form $-\mathcal{\bm{M}}(\bm{k})\cdot\bm{B}$.  They apply this to a two band model of gapped graphene and explicitly show that the phase offset ($\gamma$) in the magnetic oscillations is zero in this case even though the Berry phase around the cyclotron orbit is not $\pi$.  Because of the gap, it is instead equal to $\pi W_c[1-\Delta/\mu]$\cite{ZLi:2014a} where $W_c$ is the winding number and is a topological invariant referred to by Fuchs et al\cite{Fuchs:2010} as the topological part of the Berry phase.  They establish that $\gamma=1/2-W_c/2=0$ in this case and it is only the topological part of the Berry phase which enters this quantity.  Thus, the $-W_c/2$ exactly cancels the Maslov index contribution of 1/2\cite{Fuchs:2010}.  Following this line of reasoning, Wright and McKenzie\cite{Wright:2013} extend this semiclassical method to the case when a Schr{\"o}dinger mass is introduced as well as a gap.  In this case, the result for the phase offset is given by their Eqn.~(26).  To lowest order in the Schr{\"o}dinger mass, this reduces to $\Delta/(2mv_F^2)$.  We derive this result without resorting to any semiclassical arguments, and further find its generalization to include the Zeeman term.

\section{Surface State Hamiltonian}

\begin{figure}[h!]
\begin{center}
\includegraphics[width=1.0\linewidth]{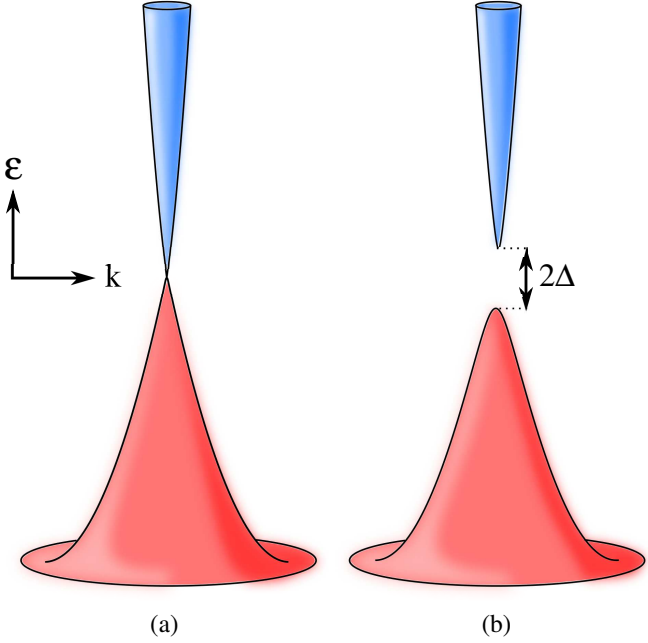}
\end{center}
\caption{\label{fig:Energy}(Color online) (a) Gapless spectrum of topological surface states. (b) The effect of including magnetic dopants on the low-energy band structure.  In both cases, significant particle-hole asymmetry is observed as the mass term causes the conduction band (blue) to narrow while the valence band (red) fans out.
}
\end{figure}

In the presence of magnetic dopants, the helical surface states of a TI are given by the Bychkov-Rashba Hamiltonian\cite{Bychkov:1984a, Bychkov:1984}
\begin{align}\label{HAM}
\hat{H}=\frac{\hbar^2 k^2}{2m}+\hbar v_F (k_x\hat{\sigma}_y-k_y\hat{\sigma}_x)+\Delta\hat{\sigma}_z,
\end{align}
where $\hat{\sigma}_x$, $\hat{\sigma}_y$ and $\hat{\sigma}_z$ are the usual Pauli-spin matrices and $\hbar\bm{k}$ is the momentum measured relative to the $\Gamma$-point of the surface Brillouin zone.  The first term is the Schr{\"o}dinger piece for describing electrons with effective mass $m$.  The linear-in-momentum term describes massless Dirac fermions which move with a Fermi velocity $v_F$.  The last term accounts for the magnetic impurities which open a gap of 2$\Delta$ in the band structure.   For Bi$_2$Te$_3$, band structure calculations\cite{Zhang:2009,Liu:2010} predict: $v_F=4.3\times 10^5$m/s, and $m=0.09m_e$ where $m_e$ is the bare mass of an electron\cite{ZLi:2014}.  Solving Eqn.~\eqref{HAM} yields the energy dispersion
\begin{align}
E_\pm(\bm{k})=\frac{\hbar^2k^2}{2m}\pm\sqrt{(\hbar v_F k)^2+\Delta^2}.
\end{align}
A schematic plot of the surface-state band structure is given in Fig.~\ref{fig:Energy}. 
Figure~\ref{fig:Energy}(a) shows the gapless spectrum which we contrast to frame (b) where a finite $\Delta$ is included.  It is important to note that we are mapping a continuum Hamiltonian onto a finite Brillouin zone.  Equation~\ref{HAM} permits the valence band to bend back toward the zero energy axis; this is unphysical for TIs.  As a result, an appropriate momentum cutoff must be applied.

To examine the magnetic response of such systems\cite{Tabert:2013a, Tabert:2013c, Tabert:2015a, ZLi:2013, Gusynin:2007, Gusynin:2006a, Tse:2010, Scharf:2013, Koshino:2008, Abergel:2007, Tkachov:2011, Tkachov:2011a}, we must consider the effect of a finite magnetic field $B$ oriented perpendicular to the surface ($\hat{z}$).  In the Landau gauge, the magnetic vector potential, defined through $\bm{B}=\nabla\times\bm{A}$, may be written as $\bm{A}=(0,Bx,0)$.  We account for the magnetic field by applying the usual Peierls substitution to the momentum ($\hbar\bm{k}\rightarrow\hbar\bm{k}+e\bm{A}$) in Eqn.~\eqref{HAM}.  Our Hamiltonian becomes
\begin{align}\label{HAM-B}
 \hat{H}&=\frac{\hbar^2}{2m}\left[(-i\partial_x)^2+(-i\partial_y+eBx/\hbar)^2\right]+\Delta\hat{\sigma}_z\notag\\
 & +\hbar v_F[(-i\partial_x)\hat{\sigma}_y-(-i\partial_y+eBx/\hbar)\hat{\sigma}_x].
\end{align}
A Zeeman interaction can be included by adding $-(1/2)g_s\mu_BB\hat{\sigma}_z$, where $g_s$ is the coupling strength ($\approx 8$ for Bi$_2$Se$_3$\cite{Wang:2010}) and $\mu_B=e\hbar/(2m_e)\approx 5.78\times 10^{-2}$ meV/T is the Bohr magneton.  Next, we define the raising and lowering operators
\begin{align}
a^\dagger\equiv\frac{l_B}{\sqrt{2}}\left[-ik_x+\frac{x+x_0}{l_B^2}\right],
\end{align}
and
\begin{align}
a\equiv\frac{l_B}{\sqrt{2}}\left[ik_x+\frac{x+x_0}{l_B^2}\right],
\end{align}
where $l_B\equiv\sqrt{\hbar/(eB)}$ and $x_0\equiv k_yl_B^2$.  On a Fock state $(\left|N\right\rangle)$ of the harmonic oscillator Hamiltonian, these operators have the property:
\begin{align}\label{a}
a\left|N\right\rangle=\sqrt{N}\left|N-1\right\rangle,
\end{align}
and
\begin{align}\label{adag}
a^\dagger\left|N\right\rangle=\sqrt{N+1}\left|N+1\right\rangle.
\end{align}
With these definitions, the Hamiltonian reduces to
\begin{align}\label{HAM-Matrix}
\hat{H}&=\left(\begin{array}{cc}
\displaystyle \frac{\hbar^2}{ml_B^2}\left[a^\dagger a+\frac{1}{2}\right]+\mathcal{Z} & \displaystyle -\hbar v_F\frac{\sqrt{2}}{l_B} a\\
\displaystyle -\hbar v_F\frac{\sqrt{2}}{l_B} a^\dagger & \displaystyle \frac{\hbar^2}{ml_B^2}\left[a^\dagger a+\frac{1}{2}\right]-\mathcal{Z}
\end{array}\right),
\end{align}
where $\mathcal{Z}\equiv\Delta-g_s\mu_BB/2$.  Equation~\eqref{HAM-Matrix} yields the LL dispersion
\begin{align}\label{LL-TI}
\mathcal{E}_{N,s}=\left\lbrace\begin{array}{cc}
\displaystyle E_0N+s\sqrt{2NE_1^2+\left[\Delta-\frac{E_0}{2}(1+g)\right]^2} & N=1,2,3,...\\
\displaystyle\frac{E_0}{2}(1+g)-\Delta & N=0
\end{array}\right.,
\end{align}
where $E_1\equiv\hbar v_F\sqrt{eB/\hbar}$, $E_0\equiv\hbar eB/m$, $g=g_sm/(2m_e)$ is the renormalized Zeeman coupling coefficient, and $s=\pm$ for the conduction and valence band, respectively. It is important to note that in a sum over $s$, the $N=0$ level will only contribute once.  Unless otherwise noted, we will take $E_1/\sqrt{B}=10.4$ meV$/\sqrt{\rm T}$, $E_0/B=1.1$ meV/T and $B=1$T.  We note that Eqn.~\eqref{LL-TI} is identical to that given previously by Shen et al.\cite{Shen:2004,Shen:2005}.  However, these authors are only interested in the limit of spintronic materials for which the Schr{\"o}dinger term in Eqn.~\eqref{HAM} is dominant.  Here, we treat the opposite limit where the Dirac term is dominant (requiring a band cutoff).  Thus, there is no overlap between the two studies and the works are complimentary.  

Since our Hamiltonian is written in the spin basis, the eigenstates are comprised of spin-up and -down amplitudes.  Indeed, the wave-functions of Eqn.~\eqref{HAM-Matrix} are
\begin{align}
\left|Ns\right\rangle=\left(\begin{array}{c}
\mathcal{C}^\uparrow_{N,s}\left|N-1\right\rangle\\
\mathcal{C}^\downarrow_{N,s}\left|N\right\rangle
\end{array}\right),
\end{align}
where
\begin{align}\label{Cup}
\mathcal{C}^\uparrow_{N,s}=\left\lbrace\begin{array}{cc}
\displaystyle -s\sqrt{\frac{1}{2}+s\frac{\Delta-(1+g)E_0/2}{2(\mathcal{E}_{N,+}-E_0N)}} & N=1,2,3,...\\
0 & N=0
\end{array}\right.,
\end{align}
is the spin-up component and
\begin{align}\label{Cdown}
\mathcal{C}^\downarrow_{N,s}=\left\lbrace\begin{array}{cc}
\displaystyle\sqrt{\frac{1}{2}-s\frac{\Delta-(1+g)E_0/2}{2(\mathcal{E}_{N,+}-E_0N)}} & N=1,2,3,...\\
1 & N=0
\end{array}\right.,
\end{align}
gives the spin-down amplitude.  We immediately see that the $N=0$ LL is entirely populated by spin-down electrons\cite{Vazifeh:2012} (i.e. $C_{N,s}^\uparrow=0$).  To see the $z$ component of spin of the remaining levels, we compute the average
\begin{align}
\left\langle \hat{s}_z\right\rangle=\left\langle N\right|\frac{\hbar}{2}\hat{\sigma}_z\left|N\right\rangle.
\end{align}  
This gives
\begin{align}
\left\langle \hat{s}_z\right\rangle=\frac{\hbar}{2}\left\lbrace\begin{array}{cc}
\displaystyle \frac{s\left[\Delta-(1+g)E_0/2\right]}{\sqrt{2E_1^2N+\left[\Delta-(1+g)E_0/2\right]^2}} & N\neq 0\\
-1 & N=0
\end{array}\right..
\end{align}
Clearly, an $\hat{s}_z$ spin polarization is quickly lost for $N\neq 0$.

\begin{figure}[h!]
\begin{center}
\includegraphics[width=1.0\linewidth]{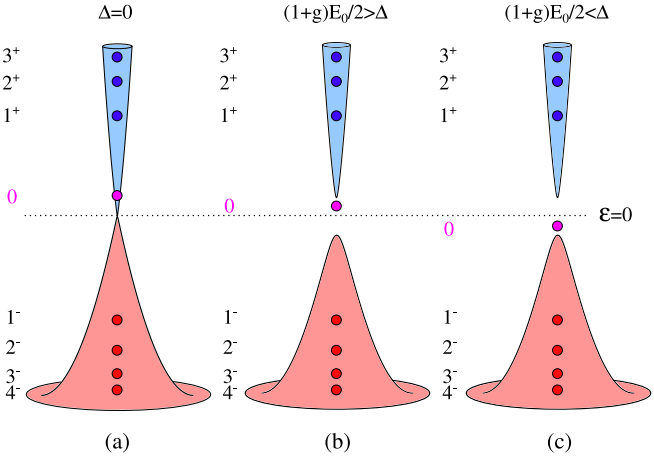}
\end{center}
\caption{\label{fig:LL-TI}(Color online) Schematic illustration of the Landau level energies for a TI with (a) $\Delta=0$, (b) $(1+g)E_0/2>\Delta$ and (c) $\Delta>(1+g)E_0/2$.  As $\Delta$ increases, the zeroth level begins at positive energy and decreases (becoming negative when $\Delta>(1+g)E_0/2$).
}
\end{figure}

Examining the $N=0$ level [Eqn.~\eqref{LL-TI}] reveals that for $\Delta=0$, the LL sits at positive energy (i.e. $\mathcal{E}_{0,+}=(1+g)E_0/2$).  For finite $\Delta$, the level remains above zero as long as $(1+g)E_0/2>\Delta$.  For $\Delta>(1+g)E_0/2$, the zeroth level is situated at negative energy.  Also, in contrast to graphene (where spin is a good quantum number), a finite Zeeman term does not split the levels but simply renormalizes their energy.  As opposed to graphene, the $N\neq 0$ levels are not simply shifted by $\pm g_s\mu_B B/2$.  The zeroth level is translated in the usual way since it has a definite spin-down polarization.  A schematic plot of the low-energy LLs in a TI is shown in Fig.~\ref{fig:LL-TI}.  
Particular attention is given to the relative location of the $N=0$ level.  The various LL energies are given by the coloured circles (blue for conduction levels, red for valence levels, and purple for $N=0$) and are overlaid on the $B=0$ band structure.

Throughout this paper, we will be interested in comparing the results of a TI with the familiar case of gapped graphene; in particular, the graphene result at the $K$-point.  In the presence of an asymmetry gap $\Delta$, the graphene Hamiltonian written about a single valley $\xi$ in the sublattice basis becomes
\begin{align}\label{Ham-G}
\hat{H}_{\xi}=\left(\begin{array}{cc}
\Delta & \hbar v(\xi k_x-ik_y)\\
\hbar v(\xi k_x+ik_y) & -\Delta
\end{array}\right),
\end{align}
where $\xi=\pm$ for the $K$ and $K^\prime$ points of the hexagonal Brillouin zone.  Applying the same techniques as before, the LL energies are\cite{Tabert:2013a, Tabert:2013c, Tabert:2015a}
\begin{align}\label{LL-G}
\mathcal{E}^{\xi\sigma}_{N,s}=\left\lbrace\begin{array}{cc}
-\frac{1}{2}g_s\mu_BB\sigma+s\sqrt{\Delta^2+2NE_1^2}, & N=1,2,3,...\\
-\xi\Delta-\frac{1}{2}g_s\mu_BB\sigma, & N=0
\end{array}\right.,
\end{align}
where $\sigma=\pm$ for spin up and down, respectively.  The corresponding wave-functions are
\begin{equation}
\left|N s\right\rangle_K=
\left(\begin{array}{c}
-i\mathcal{A}_{N,s}\left|N-1\right\rangle\\
\mathcal{B}_{N,s}\left|N\right\rangle
\end{array}\right)
\end{equation}
and
\begin{equation}
\left|Ns\right\rangle_{K^\prime}=
\left(\begin{array}{c}
-i\mathcal{A}_{N,s}\left|N\right\rangle\\
\mathcal{B}_{N,s}\left|N-1\right\rangle
\end{array}\right),
\end{equation}
respectively, where\cite{Tabert:2013a, Tabert:2013c, Tabert:2015a}
\begin{equation}\label{An}
\mathcal{A}_{N,s}=\left\lbrace\begin{array}{cc}
\displaystyle\frac{s\sqrt{|\mathcal{E}_{N,s}^{\xi\sigma}|+s\Delta}}{\sqrt{2|\mathcal{E}_{N,s}^{\xi\sigma}|}}, &\quad N\neq 0,\\
\displaystyle\frac{1-\xi}{2}, &\quad N=0,
\end{array}\right.
\end{equation}
and
\begin{equation}\label{Bn}
\mathcal{B}_{N,s}=\left\lbrace\begin{array}{cc}
\displaystyle\frac{\sqrt{|\mathcal{E}_{N,s}^{\xi\sigma}|-s\Delta}}{\sqrt{2|\mathcal{E}_{N,s}^{\xi\sigma}|}}, &\quad N\neq 0,\\
\displaystyle\frac{1+\xi}{2}, &\quad N=0.
\end{array}\right.
\end{equation}
Unlike the TI, spin is a good quantum number and thus a Zeeman interaction simply shifts the spin-up (-down) levels down (up) by a constant amount.

\section{Density of States}

\begin{figure}[h!]
\begin{center}
\includegraphics[width=1.0\linewidth]{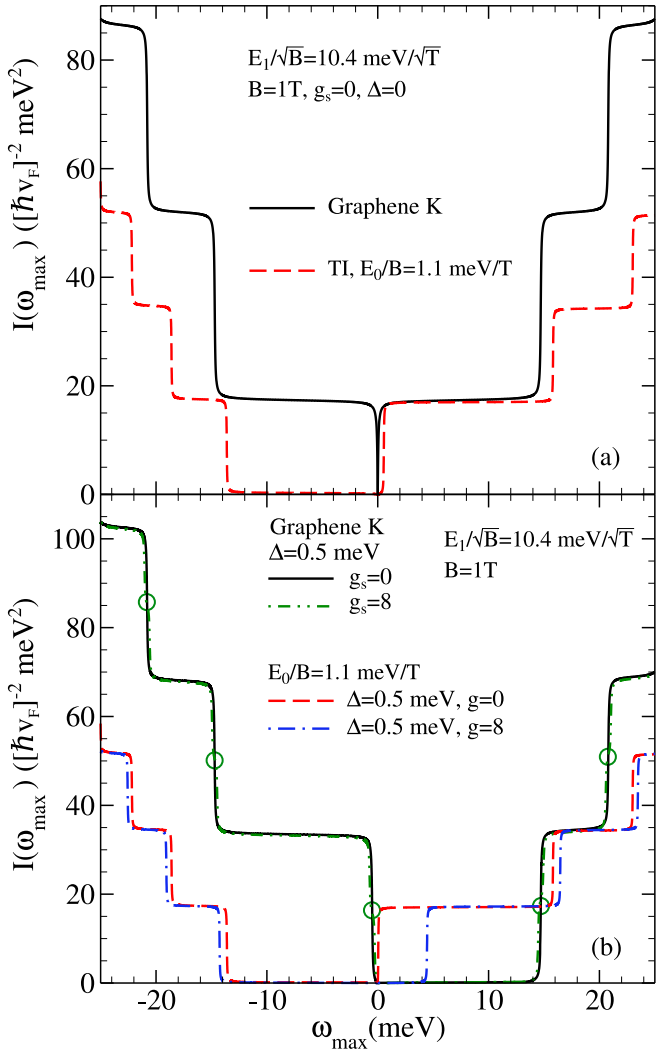}
\end{center}
\caption{\label{fig:DOS-Sum}(Color online) Integrated density of states for graphene at the $K$-point and a TI (a) without and (b) with a gap of $\Delta=0.5$ meV.  (a) For $\Delta=0$, particle-hole symmetry is present in graphene.  For a TI the valence and conduction levels are asymmetric.  (b) For finite $\Delta$, the symmetry in graphene is removed.  A Zeeman interaction spin-splits the graphene levels and simply renormalized those of a TI.
}
\end{figure}

\begin{figure}[h!]
\begin{center}
\includegraphics[width=1.0\linewidth]{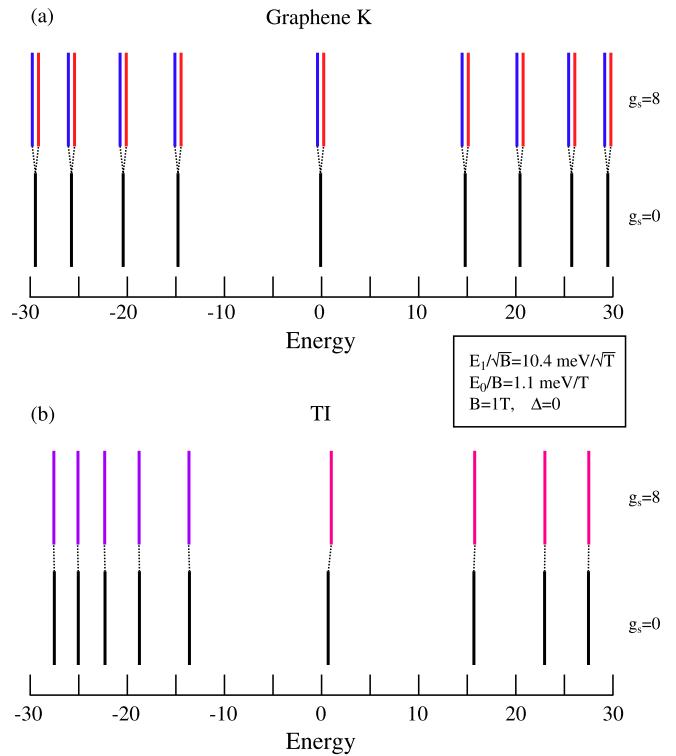}
\end{center}
\caption{\label{fig:DOS}(Color online) $\Delta=0$ LLs in (a) graphene and (b) a TI with and without Zeeman splitting. (a) In graphene, a Zeeman interaction spin-splits all the LLs and shifts the spin-down levels (red) up while the spin-up LLs (blue) decrease in energy.  (b) For a TI, Zeeman effects do not split the levels.
}
\end{figure}

We begin our analysis with a comparison of the integrated density of states.  In a magnetic field, the density of states is given by a sum of $\delta$-functions which peak at the various LL energies.  For a TI,
\begin{align}\label{DOS}
N(\omega)=\frac{eB}{h}\left[\delta\left(\omega-\mathcal{E}_{0,+}\right)+\sum_{\substack{N=1 \\ s=\pm}}^\infty\delta\left(\omega-\mathcal{E}_{N,s}\right)\right],
\end{align}
where $\mathcal{E}_{N,s}$ is given by Eqn.~\eqref{LL-TI}.  Similarly, for graphene
\begin{align}
N_{\xi\sigma}(\omega)=\frac{eB}{h}\left[\delta\left(\omega-\mathcal{E}^{\xi\sigma}_{0,+}\right)+\sum_{\substack{N=1 \\ s=\pm}}^\infty\delta\left(\omega-\mathcal{E}^{\xi\sigma}_{N,s}\right)\right],
\end{align}
where $\mathcal{E}^{\xi\sigma}_{N,s}$ is given by Eqn.~\eqref{LL-G}.  To compare the two results, we define the total integrated density of states up to energy $\omega_{\rm max}$ as
\begin{align}\label{DOS-Int}
I(\omega_{\rm max})=\int_0^{\omega_{\rm max}}N(\omega)d\omega.
\end{align}
A plot of $I(\omega_{\rm max})$ as function of the cutoff energy $\omega_{\rm max}$ is shown in Fig.~\ref{fig:DOS-Sum}.  For graphene, the steps are of double weight due to spin-degeneracy. 
Figure~\ref{fig:DOS-Sum}(a) shows the $\Delta=0$ results for graphene at the $K$-point (solid black) and a TI (dashed red) in the absence of Zeeman splitting.  For graphene, the $N=0$ LL is pinned at zero energy and particle-hole symmetry is present for all $N$.  For the TI, the $N=0$ level sits at positive energy and clear particle-hole asymmetry exists for the $N\neq 0$ levels.  The valence levels sit closer to $\omega_{\rm max}=0$ than the corresponding conduction levels.  In Fig.~\ref{fig:DOS-Sum}(b), a finite gap is included and the effect of Zeeman splitting is explored.  For graphene with $g_s=0$ (solid black), the gap moves the $N=0$ level to negative energy at the $K$-point and breaks the particle-hole symmetry.  With a finite Zeeman term (dash-double-dotted green), the steps split into two (emphasized by the green circles).  This is characteristic of systems with spin-degeneracy.  For a TI (dashed red), the gap shifts the energy levels and a finite Zeeman interaction (dash-dotted blue) does not split the steps but further renormalizes their energy.  A plot of the LLs for $\Delta=0$ and varying $g_s$ is shown in Fig.~\ref{fig:DOS}.  
For graphene at $K$ [frame (a)], the $g_s=0$ levels are symmetric.  A finite Zeeman splitting causes the spin-down levels (red) to increase in energy by $g_s\mu_BB/2$ while the spin-up levels (blue) decrease by the same amount.  This splitting does not depend on $N$ and creates a series of spin-polarized levels.  The TI [frame (b)] is particle-hole asymmetric and a Zeeman term causes the conduction levels (pink) to increase in energy while the valence levels (purple) decrease.  Unlike graphene, the $N\neq 0$ levels are not spin-polarized and the splitting is not uniform; for larger $N$, the renormalization becomes less prominent.

\section{Magneto-Optical Conductivity}

In the one-loop approximation, the optical conductivity $\sigma_{\alpha\beta}(\Omega)$ is given by the familiar Kubo formula
\begin{align}\label{Kubo-Mag}
\sigma_{\alpha\beta}(\Omega)=\frac{i\hbar}{2\pi l_B^2}&\sum_{\substack{N,M=0 \\s,s^\prime=\pm}}^\infty\frac{f_{M,s^\prime}-f_{N,s}}{\mathcal{E}_{N,s}-\mathcal{E}_{M,s^\prime}}\notag\\
&\times\frac{\left\langle \bar{N}s|\hat{j}_\alpha|\bar{M}s^\prime\right\rangle\left\langle \bar{M}s^\prime|\hat{j}_\beta|\bar{N}s\right\rangle}{\hbar\Omega+\mathcal{E}_{M,s^\prime}-\mathcal{E}_{N,s}+i\hbar/(2\tau)},
\end{align}
where $f_{n,s}$ is the Fermi function for state $n$ in band $s$, $\tau$ is the relaxation time and $\hat{j}_{\alpha}=e\hat{v}_\alpha\equiv (e/\hbar)(\partial \hat{H}/\partial k_{\alpha})$ is the current operator.  In the zero temperature limit, $f_n$ can be replaced by the Heaviside step function $\Theta(\mu-\mathcal{E}_{n,s})$.  The necessary velocity operators are
\begin{align}\label{vx}
\hat{v}_x=\frac{\hbar}{m}k_x+v_F\hat{\sigma}_y=i\frac{\hbar}{m}\frac{a^\dagger-a}{\sqrt{2}l_B}+v_F\hat{\sigma}_y,
\end{align}
and
\begin{align}\label{vy}
\hat{v}_y=\frac{\hbar}{m}\left[k_y+\frac{eBx}{\hbar}\right]-v_F\hat{\sigma}_x=\frac{\hbar}{m}\frac{a^\dagger+a}{\sqrt{2}l_B}-v_F\hat{\sigma}_x.
\end{align}
Calculating the appropriate matrix elements, we obtain the real and imaginary parts of the longitudinal conductivity
\begin{align}\label{sigmaxx}
{\rm Re}\left\lbrace\frac{\sigma_{xx}(\Omega)}{e^2/\hbar}\right\rbrace &=\frac{E_1^2}{2\pi}\sum_{\substack{N,M=0 \\s,s^\prime=\pm}}^\infty\frac{f_{M,s^\prime}-f_{N,s}}{\mathcal{E}_{N,s}-\mathcal{E}_{M,s^\prime}}\\
&\times\frac{\Gamma}{(\Omega+\mathcal{E}_{M,s^\prime}-\mathcal{E}_{N,s})^2+\Gamma^2}\notag\\
&\times[\mathcal{F}(Ns;Ms^\prime)\delta_{N,M-1}+\mathcal{F}(Ms^\prime;Ns)\delta_{M,N-1}],\notag
\end{align}
and
\begin{align}\label{sigmaxx-i}
{\rm Im}\left\lbrace\frac{\sigma_{xx}(\Omega)}{e^2/\hbar}\right\rbrace &=\frac{E_1^2}{2\pi}\sum_{\substack{N,M=0 \\s,s^\prime=\pm}}^\infty\frac{f_{M,s^\prime}-f_{N,s}}{\mathcal{E}_{N,s}-\mathcal{E}_{M,s^\prime}}\\
&\times\frac{\Omega+\mathcal{E}_{M,s^\prime}-\mathcal{E}_{N,s}}{(\Omega+\mathcal{E}_{M,s^\prime}-\mathcal{E}_{N,s})^2+\Gamma^2}\notag\\
&\times[\mathcal{F}(Ns;Ms^\prime)\delta_{N,M-1}+\mathcal{F}(Ms^\prime; Ns)\delta_{M,N-1}],\notag
\end{align}
respectively, where $\Gamma\equiv\hbar/(2\tau)$ and
\begin{align}
\mathcal{F}(Ns;Ms^\prime)&\equiv\left[\mathcal{C}^\uparrow_{M,s^\prime}\mathcal{C}^\downarrow_{N,s}-\frac{E_0}{\sqrt{2}E_1}\right.\\
&\times\left.\left(\sqrt{N}\mathcal{C}^\uparrow_{M,s^\prime}\mathcal{C}^\uparrow_{N,s}+\sqrt{N+1}\mathcal{C}^\downarrow_{M,s^\prime}\mathcal{C}^\downarrow_{N,s}\right)\right]^2.\notag
\end{align}
Likewise, the real and imaginary parts of the transverse Hall conductivity are
\begin{align}\label{sigmaxy}
{\rm Re}\left\lbrace\frac{\sigma_{xy}(\Omega)}{e^2/\hbar}\right\rbrace &=\frac{E_1^2}{2\pi}\sum_{\substack{N,M=0 \\s,s^\prime=\pm}}^\infty\frac{f_{M,s^\prime}-f_{N,s}}{\mathcal{E}_{N,s}-\mathcal{E}_{M,s^\prime}}\\
&\times\frac{\Omega+\mathcal{E}_{M,s^\prime}-\mathcal{E}_{N,s}}{(\Omega+\mathcal{E}_{M,s^\prime}-\mathcal{E}_{N,s})^2+\Gamma^2}\notag\\
&\times[\mathcal{F}(Ns;Ms^\prime)\delta_{N,M-1}-\mathcal{F}(Ms^\prime; Ns)\delta_{M,N-1}],\notag
\end{align}
and
\begin{align}\label{sigmaxy-i}
{\rm Im}\left\lbrace\frac{\sigma_{xy}(\Omega)}{e^2/\hbar}\right\rbrace &=\frac{E_1^2}{2\pi}\sum_{\substack{N,M=0 \\s,s^\prime=\pm}}^\infty\frac{f_{M,s^\prime}-f_{N,s}}{\mathcal{E}_{N,s}-\mathcal{E}_{M,s^\prime}}\\
\times &\frac{-\Gamma}{(\Omega+\mathcal{E}_{M,s^\prime}-\mathcal{E}_{N,s})^2+\Gamma^2}\notag\\
\times &[\mathcal{F}(Ns;Ms^\prime)\delta_{N,M-1}-\mathcal{F}(Ms^\prime;Ns)\delta_{M,N-1}].\notag
\end{align}
For the gapped graphene results, the reader is referred to Eqns. (12)-(15) in Ref.~\cite{Tabert:2013c}.  Note that an overall minus sign should be included in Eqns. (14) and (15) of Ref.~\cite{Tabert:2013c}.  To obtain the appropriate limit, $\Delta_{\rm so}=0$ and $\Delta_z/2=\Delta$.  As previously mentioned, a momentum cutoff must be applied to the band structure to ensure that the valence band does not bend back across the zero energy axis.  When a magnetic field is applied, a corresponding cutoff on $N$ must be applied to ensure that $\mathcal{E}_{N,-}<0$.

\begin{figure}[h!]
\begin{center}
\includegraphics[width=1.0\linewidth]{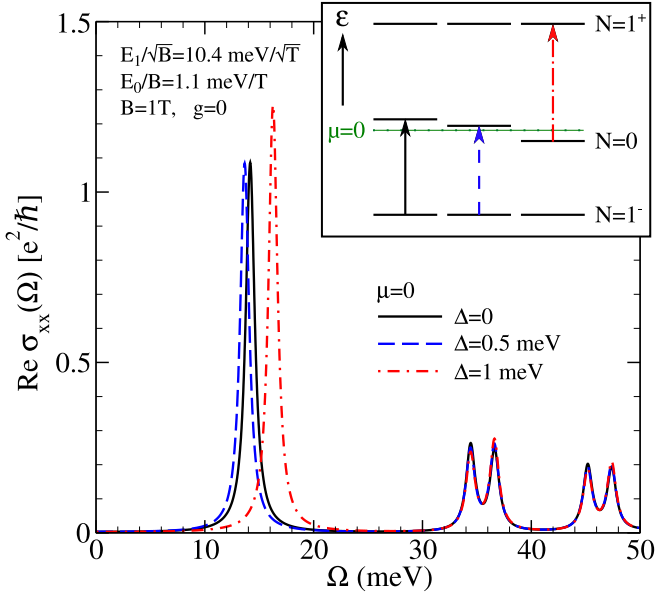}
\end{center}
\caption{\label{fig:Cond-Gap}(Color online) $\mu=0$ longitudinal conductivity in a TI for varying $\Delta$.  The energy of the first absorption process is schematically illustrated by the arrows in the inset.  As $\Delta$ increases, the zeroth LL decreases in energy. This shift is evident in the onset frequency of the first optical transition.
}
\end{figure}

We begin our discussion by consider the effect of finite $\Delta$ on the absorptive part of the longitudinal response.  This is shown in Fig.~\ref{fig:Cond-Gap} for a TI.
Re$\sigma_{xx}(\Omega)$ is plotted for $\mu=0$, $g=0$ and three values of $\Delta$: 0 (solid black), 0.5 meV (dashed blue) and 1 meV (dash-dotted red).  The inset shows a schematic plot of the lowest LLs for the different $\Delta$'s.  The lowest optical transition is marked by the arrows.  For $\Delta=0$, the $N=0$ LL sits at $E_0/2$ (for $g=0$).  Therefore, the lowest transition occurs between the $N=1$, $s=-$ and N=0 levels.  For finite $\Delta$ less than $E_0/2$, the $N=0$ level is still at positive energy but its magnitude has decreased; this causes the lowest absorption peak to move down in $\Omega$.  As $\Delta$ becomes larger than $E_0/2$, the $N=0$ level moves to negative energy and the first transition is now between the zeroth level and the $N=1$ LL of the conduction band.  Now, the frequency of the first transition continues to increase with $\Delta$ as $\mathcal{E}_{0,+}$ is pushed further down in energy.  Higher optical transitions are present and occur in pairs.  This is a signature of the particle-hole asymmetry as the energy for the $\mathcal{E}_{N,-}$ to $\mathcal{E}_{N\pm 1,+}$ transition is not the same as the $\mathcal{E}_{N\pm 1,-}$ to $\mathcal{E}_{N,+}$ transition.  For graphene, the two sets of split higher-energy peaks in Fig.~\ref{fig:Cond-Gap} would each coalesce into a single line.  It is the presence of the Schr{\"o}dinger magnetic energy $E_0$ which sets the scale for the splitting.  This energy increases linearly with magnetic field and inversely with decreasing $m$ as discussed by Li and Carbotte\cite{ZLi:2013}.

\begin{figure}[h!]
\begin{center}
\includegraphics[width=0.90\linewidth]{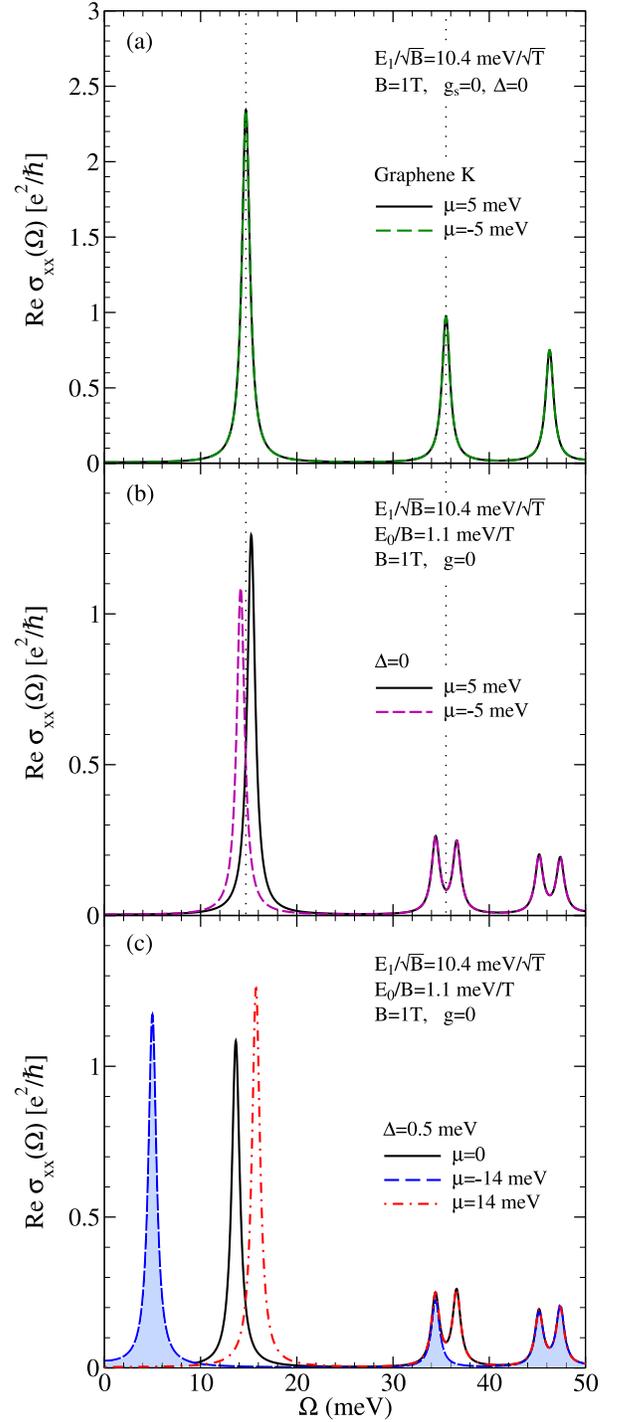}
\end{center}
\caption{\label{fig:Cond-TI-G}(Color online) Longitudinal conductivity in (a) gapless graphene, and (b) a gapless and (c) gapped TI for positive and negative $\mu$.  (a) For graphene, particle-hole symmetry ensures the $\pm \mu$ results are identical.  For a TI [(b)-(c)], the $\pm\mu$ results are different which emphasizes the asymmetry of the LLs.  The shading under the dashed blue curve is for emphasis.  Note the missing peak in this case. 
}
\end{figure}

\begin{figure}[h!]
\begin{center}
\includegraphics[width=1.0\linewidth]{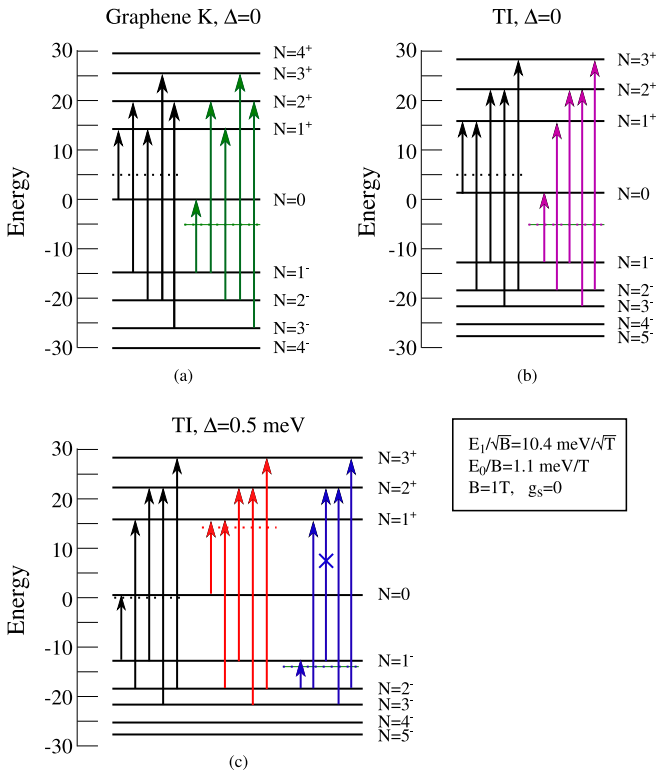}
\end{center}
\caption{\label{fig:Cond-levels}(Color online) Absorption processes which give rise to the conductivity curves of Fig.~\ref{fig:Cond-TI-G}.  The arrows are coloured to correspond with the curves in Fig.~\ref{fig:Cond-TI-G} and the location of $\mu$ is given by the dotted lines.  For $\mu=-14$ meV in the gapped TI [blue arrows in frame (c)], the transition from $\mathcal{E}_{1,-}$ to $\mathcal{E}_{2,+}$ is forbidden [note the missing absorption line in Fig.~\ref{fig:Cond-TI-G}(c)].
}
\end{figure}

Next, we explore the results for finite chemical potential and $g=0$.  In Fig.~\ref{fig:Cond-TI-G}, the effect of positive and negative $\mu$ is examined.  
In frame (a), the results of gapless graphene at the $K$-point are shown; due to the particle-hole symmetry, the response is identical for $\pm\mu$.  The optical transitions which lead to this set of absorption lines are shown in Fig.~\ref{fig:Cond-levels}(a) where the arrows are color-coded to correspond to Fig.~\ref{fig:Cond-TI-G}(a).  Note that the positions of the $\mathcal{E}_{N,-}$ LLs are at the negative of the $\mathcal{E}_{N,+}$ levels.  For positive $\mu$, the five arrows on the left (black) apply; while, for negative $\mu$, it is the five green arrows on the right which are relevant.  In both cases, each transition has a one-to-one correspondence with the other set.  Figure~\ref{fig:Cond-TI-G}(b), displays the results for a TI when $\Delta=0$.  Here the finite Schr{\"o}dinger term breaks particle-hole symmetry.  This is clear from the transitions shown in Fig.~\ref{fig:Cond-levels}(b).  In this case, even for $\Delta=0$, the negative energy levels do not mirror the positive energy set.  The $N=0$ level is no longer at zero energy but rather has been pushed to positive energy $E_0/2$.  The $\mathcal{E}_{N,+}$ energy is also larger than $|\mathcal{E}_{N,-}|$.  The black arrow between $N=0$ and $N=1^+$ (which applies to the lowest line for positive $\mu$) is longer by $E_0/2$ than the arrows between the $N=1^-$ and $N=0$ levels which is the first transition when $\mu$ is negative.  Thus, in Fig.~\ref{fig:Cond-TI-G}(b), the first peak of the negative $\mu$ response (purple) is lower in $\Omega$ than the corresponding positive $\mu$ peak (black).  The results for a gapped TI are shown in frame (c) of Fig.~\ref{fig:Cond-TI-G} with the corresponding optical transitions shown in Fig.~\ref{fig:Cond-levels}(c).  Again, an obvious asymmetry is present between the $\pm\mu$ regimes.  The asymmetry is now much larger than that shown in frame (b) for two reasons: the finite gap adds asymmetry; the larger value of chemical potential enhances it is well.  An additional effect which needs to be emphasized is that one of the absorption lines for $\mu<0$ (blue) is missing in the first set of split peaks shown at higher energy.  Of the four peaks, the first, third and fourth apply to all three curves ($\mu=0$ and $\mu=\pm 14$ meV), but the second is only present in the black and red curves.  The transitions which give rise to these plots are shown in Fig.~\ref{fig:Cond-levels} where the missing transition is indicated by a blue $\times$.  It cannot occur because the $N=1^-$ level is unoccupied.

\begin{figure}[h!]
\begin{center}
\includegraphics[width=1.0\linewidth]{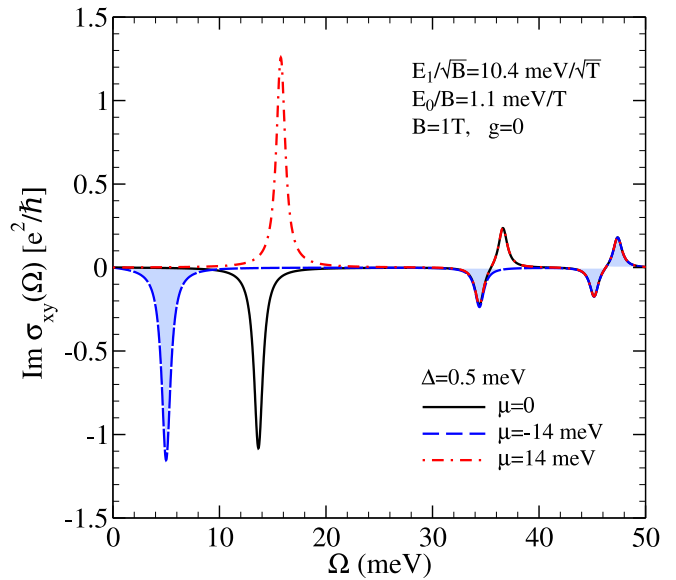}
\end{center}
\caption{\label{fig:Cond-xy}(Color online) The absorptive part of the transverse Hall conductivity for a TI with parameters set to match Fig.~\ref{fig:Cond-TI-G}(c).  For transitions from $N$ to $N-1$, the response is negative; for $N$ to $N+1$, the conductivity is positive [see Fig.~\ref{fig:Cond-levels}(c)].
}
\end{figure}

The absorptive part of the transverse Hall conductivity is also of interest and is shown in Fig.~\ref{fig:Cond-xy} for the same parameters as Fig.~\ref{fig:Cond-TI-G}(c).
When the optical transition is from the $N^{\rm th}$ level in the valence band to the $(N-1)^{\rm th}$ LL, the response is negative; while the transition from the $N^{\rm th}$ to $(N+1)^{\rm th}$ level is positive.  This will have important ramifications on the circular dichroism.  The shading under the dashed blue curve for negative $\mu$ again helps to emphasize the missing peak around $\sim 36.6$ meV which occurs for $\mu=0$ (black) and $\mu=14$ meV (red).  The other three peaks exist for all $\mu$ considered here.  Note that, expect for the lowest energy peak in the dashed curve (blue) and in the dash-dotted curve (red), all other peaks would not exist in the pure Dirac limit.  Their existence depends on the presence of the subdominant Schr{\"o}dinger contribution in our Hamiltonian, even if it is small.  This represents a qualitative difference between the physics of the two cases.

\begin{figure}[h!]
\begin{center}
\includegraphics[width=1.0\linewidth]{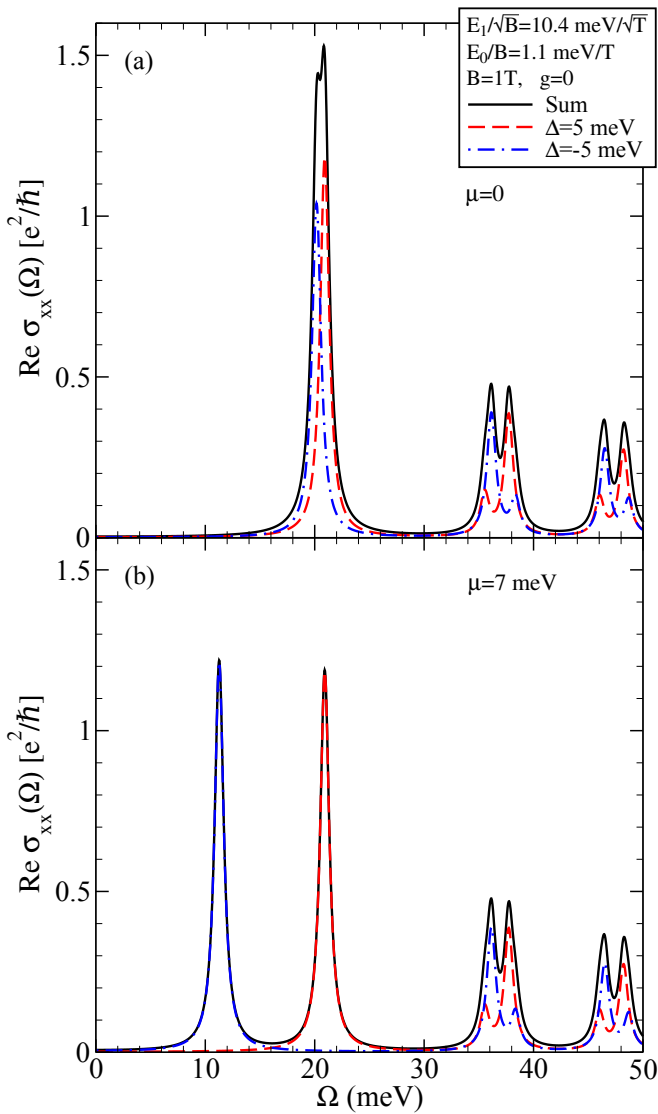}
\end{center}
\caption{\label{fig:Cond-Dpm}(Color online) The absorptive part of the longitudinal conductivity for a TI with $\Delta=\pm 5$ meV for (a) $\mu=0$ and (b) $\mu=7$ meV.
}
\end{figure}

In the context of TI thin films, both surfaces become important; the two surfaces are marked by an opposite sign of $\Delta$.  In Fig.~\ref{fig:Cond-Dpm}, the longitudinal conductivity for $\Delta=\pm 5$ meV is shown.  The combined result that would be measured in an optics experiment is given by the solid black curve.
Frame (a) corresponds to charge neutrality ($\mu=0$) while finite chemical potential is considered in frame (b).  Aside from an energy shift in the peak associated with $N=0$, we note that the spectral weight for the $N\rightarrow N\pm 1$ transitions (doublets of higher energy peaks) is markedly different depending on the sign of the gap.  In the dash-dotted blue curve, the first peak of the doublet has the largest spectral weight while it is opposite for the dashed red curve.

\subsection{Circular Dichroism}

\begin{figure}[h!]
\begin{center}
\includegraphics[width=1.0\linewidth]{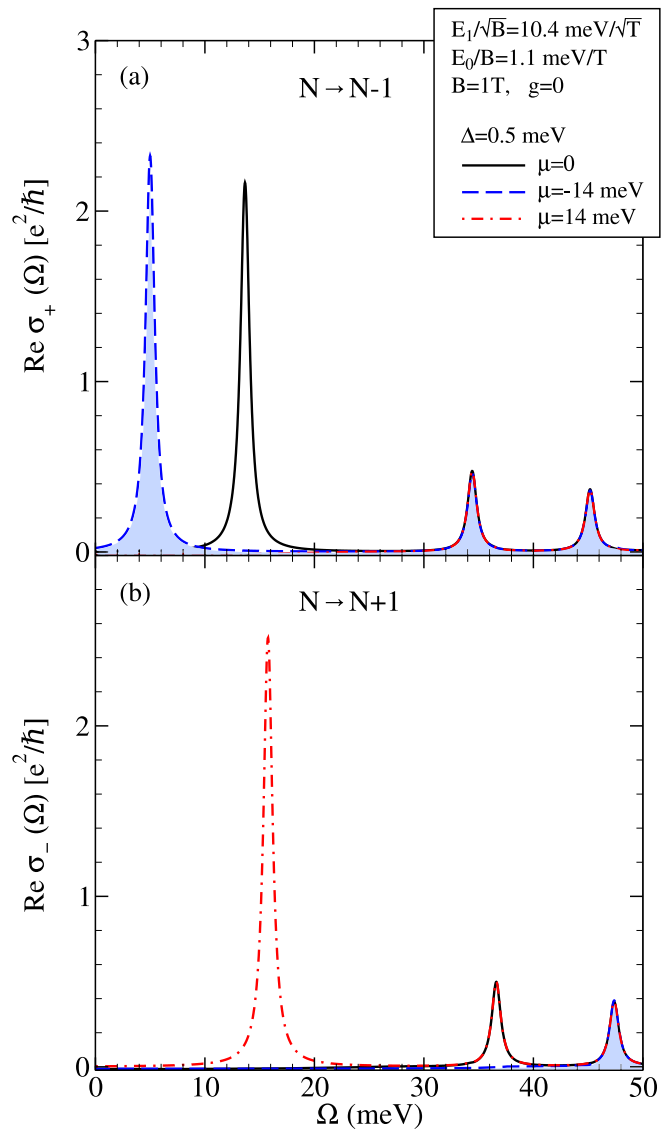}
\end{center}
\caption{\label{fig:Cond-Circ}(Color online) Conductivity response to (a) right- and (b) left-handed circularly-polarized light.  (a) For right-handed light, only transitions from $N$ to $N-1$ are present. (b) In the response to left-handed polarization, only the $N$ to $N+1$ processes occur.
}
\end{figure}

The response to circularly polarized light is given by $\sigma_{xx}(\Omega)\pm i\sigma_{xy}(\Omega)$ for right- and left-handed polarization, respectively.  Therefore, the absorptive part is determined by
\begin{align}
{\rm Re}\sigma_\pm(\Omega)={\rm Re}\sigma_{xx}(\Omega)\mp{\rm Im}\sigma_{xy}(\Omega).
\end{align}
This is readily evaluated by utilizing Eqns.~\eqref{sigmaxx} and \eqref{sigmaxy-i}. A plot of the circular dichroism is given in Fig.~\ref{fig:Cond-Circ} for a gapped TI.
Frames (a) and (b) show the response to right- and left-handed polarized light, respectively.  Right-handed light selects out the transitions between the $N^{\rm th}$ level in the valence band and the $(N-1)^{\rm th}$ LL in the conduction band.  Left-handed light selects the $N$ to $N+1$ transitions.  Right circularly polarized light shows a low-energy absorption peat at $\mu=0$ (solid black) which moves to lower energy in the dashed blue curve for $\mu=-14$ meV [frame (a)].  Such a peak is missing for $\mu=+14$ meV (dash-dotted red).  Instead, this peak is present for left circular polarization [frame (b)] while the other two values of $\mu$ have no such absorption feature.  The pairs of peaks shown at higher energy in both frames are present for all three values of $\mu$ for right-handed light.  For left-handed polarization, the higher peak in the pair is present for all $\mu$ considered here, the lower one is missing for $\mu=14$ meV (dashed blue).  This can be traced to the forbidden transition shown in Fig.~\ref{fig:Cond-levels}(c) (marked by the blue $\times$).

\section{Magnetization of the Metallic Surface States}

To discuss the magnetization of the surface states, we turn to the grand thermodynamic potential
\begin{align}\label{Omega-T}
\Omega(T,\mu)=-T\int_{-\infty}^\infty N(\omega){\rm ln}\left(1+e^{(\mu-\omega)/T}\right)d\omega,
\end{align}  
where $T$ is the temperature (in units of $k_B$), and $N(\omega)$ is given by Eqn.~\eqref{DOS}.  The quantities discussed herein, depend on a $B$ derivative of the grand potential; for convenience, we add $(\mu/2)\int_{-\infty}^\infty N(\omega)d\omega$ to Eqn.~\eqref{Omega-T}.  The integral of the density of states over all energies gives the total number of states (which must be independent of $B$).  Thus, this term will not contribute to the magnetization ($-\partial\Omega/\partial B$).  At zero temperature, Eqn.~\eqref{Omega-T} becomes
\begin{align}
\Omega(\mu)=\int_{-\infty}^0\left(\omega-\frac{\mu }{2}\right)N(\omega)d\omega &+\int_0^\mu (\omega-\mu)N(\omega)d\omega\notag\\
&+\frac{\mu}{2}\int_0^\infty N(\omega)d\omega.
\end{align}
For a gapped TI,
\begin{align}
\int_0^\infty N(\omega)d\omega=\int_{-\infty}^0 N(\omega)d\omega+\frac{eB}{h}\Upsilon,
\end{align}
where 
\begin{align}
\Upsilon=\left\lbrace\begin{array}{cc}
-1, & \displaystyle(1+g)E_0/2<\Delta\\
0, & \displaystyle(1+g)E_0/2=\Delta\\
1, & \displaystyle(1+g)E_0/2>\Delta\\
\end{array}\right..
\end{align}
Therefore, 
\begin{align}\label{Omega-minus-vac}
\Omega(\mu)=\int_0^\mu (\omega-\mu)N(\omega)d\omega+\frac{eB\mu}{2h}\Upsilon+\int_{-\infty}^0\omega N(\omega)d\omega.
\end{align}
The final term does not depend on $\mu$ and gives the vacuum contribution.  This will simply provide a constant background to the $\mu$ dependence of the magnetization.    Keeping only the $\mu$ dependent pieces, the first two terms of Eqn.~\eqref{Omega-minus-vac} give
\begin{align}\label{Omega-tilde}
\tilde{\Omega}(\mu)=\frac{eB}{h}& \bigg[{\rm sgn}(\mathcal{E}_{0,+})\left(\mathcal{E}_{0,+}-\mu\right)\Theta\left(|\mu|-|\mathcal{E}_{0,+}|\right)\Theta\left({\rm sgn}(\mathcal{E}_{0,+})\mu\right)\notag\\
&+\frac{\Upsilon\mu}{2}+\sum_{N=1}^\infty\left(\mathcal{E}_{N,+}-\mu\right)\Theta\left(\mu-\mathcal{E}_{N,+}\right)\notag\\
&-\sum_{N=1}^\infty\left(\mathcal{E}_{N,-}-\mu\right)\Theta\left(\mathcal{E}_{N,-}-\mu\right)\bigg],
\end{align}
where again, we require all the $s=-$ states to be negative.  Note that $\Theta(0)\equiv 1/2$ as only half the $\delta$-function situated at $\omega=0$ is integrated.  

The slope of the magnetization is of particular interest as it is related to the quantized Hall conductivity through the Streda relation\cite{Smrcka:1977} ($\sigma_H=e\partial M/\partial\mu$).  To see the quantization of the slope, note that
\begin{align}
\frac{\partial\Omega}{\partial\mu}=-\frac{eB}{h}& \bigg[{\rm sgn}(\mathcal{E}_{0,+})\Theta\left(|\mu|-|\mathcal{E}_{0,+}|\right)\Theta\left({\rm sgn}(\mathcal{E}_{0,+})\mu\right)\notag\\
&-\frac{\Upsilon}{2}+\sum_{N=1}^\infty\Theta\left(\mu-\mathcal{E}_{N,+}\right)-\Theta\left(\mathcal{E}_{N,-}-\mu\right)\bigg].
\end{align}
The slope of the magnetization is then
\begin{align}\label{M-dmu}
\frac{\partial M}{\partial\mu}=-&\frac{\partial}{\partial B}\frac{\partial\Omega}{\partial\mu}=\frac{e}{h}\bigg[{\rm sgn}(\mathcal{E}_{0,+})\Theta\left(|\mu|-|\mathcal{E}_{0,+}|\right)\Theta\left({\rm sgn}(\mathcal{E}_{0,+})\mu\right)\notag\\
&-\frac{\Upsilon}{2}+\sum_{N=1}^\infty\Theta\left(\mu-\mathcal{E}_{N,+}\right)-\Theta\left(\mathcal{E}_{N,-}-\mu\right)\bigg].
\end{align}

\begin{figure}[h!]
\begin{center}
\includegraphics[width=1.0\linewidth]{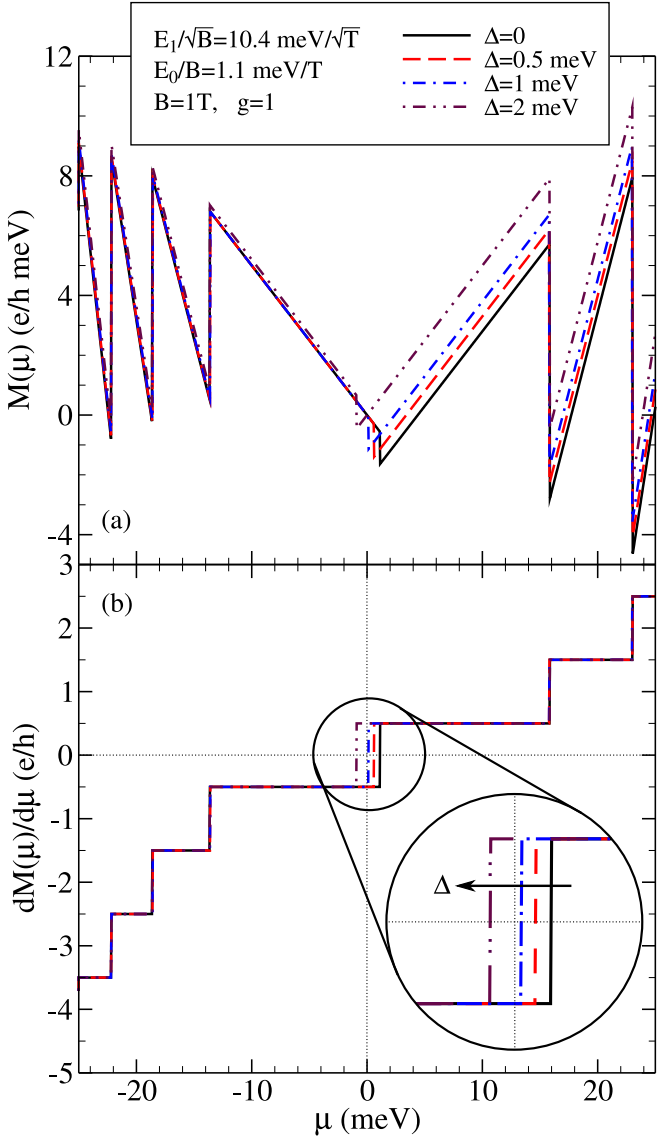}
\end{center}
\caption{\label{fig:QHE}(Color online) (a) $\mu$ dependence of the magnetization for a TI with varying $\Delta$. (b) The corresponding slope of the magnetization (which is related to the Hall conductivity through the Streda relation: $\partial M/\partial\mu=(1/e)\sigma_H$).  In all cases, a half-integer quantization is present; however, the step to $+1/2$ can be  tuned from positive to negative $\mu$ by increasing $\Delta$.
}
\end{figure}

A plot of the magnetization calculated from Eqn.~\eqref{Omega-tilde} through $M=-\partial\Omega/\partial B$ is shown in Fig.~\ref{fig:QHE}(a).  The corresponding slope [see Eqn.~\eqref{M-dmu}] is given in Fig.~\ref{fig:QHE}(b).  Here, $g$ is taken to be 1 and the effect of varying $\Delta$ is emphasized.
A saw-tooth oscillation pattern is present in $M(\mu)$ with the location of the teeth sitting at the various LL energies.  The derivative of $M(\mu)$ gives the quantization of the Hall conductivity.  A half-integer quantization is present which is characteristic of Dirac systems.  The Hall conductivity is $\sigma_H=(e^2/h)\nu$ with filling factors $\nu=\pm 1/2,\pm 3/2, \pm 5/2,...$.  For zero gap (solid black curve), the Hall conductivity at $\mu=0$ has $\nu=-1/2$.  This results from the finite value of the $N=0$ level\cite{Tabert:2015}.  As $\Delta$ is increased, the location of the $N=0$ step moves lower in $\mu$.  At $\mu=(1+g)E_0/2-\Delta$, the step occurs at zero chemical potential; for larger $\Delta$, the small $|\mu|$ value of $\sigma_H$ is given by $\nu=1/2$.  Again, asymmetry is seen between the negative and positive $\mu$ regimes.  These results have been verified by taking the DC limit ($\Omega\rightarrow 0$) of Re$\sigma_{xy}(\Omega)$ [see Eqn.~\eqref{sigmaxy}].

\begin{figure}[h!]
\begin{center}
\includegraphics[width=1.0\linewidth]{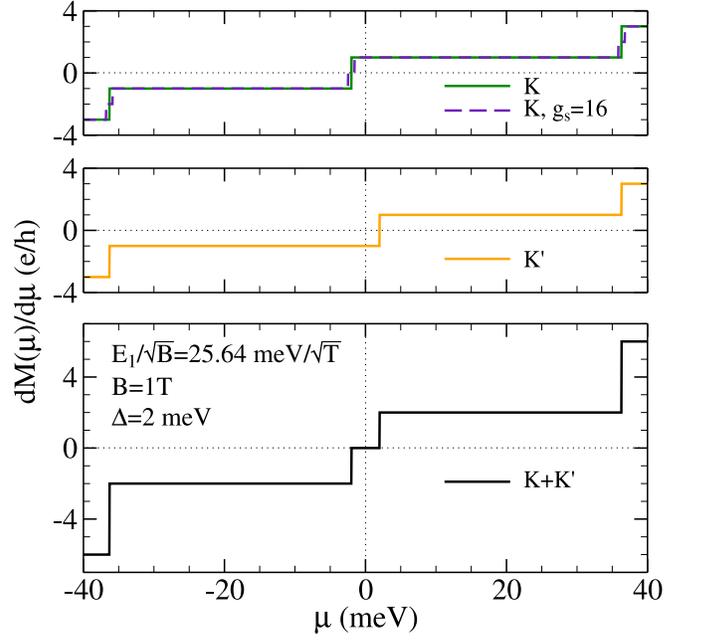}
\end{center}
\caption{\label{fig:QHE-G}(Color online) The slope of the magnetization for gapped graphene at (top) the $K$-point, (middle) the $K^\prime$-point and (lower) the total response.  The step which results from the $N=0$ level is at $\mu=\mp\Delta$ for the $K$- and $K^\prime$-point, respectively.  Thus, the total system is insulating for $-\Delta<\mu<\Delta$. Zeeman (dashed purple curve in the top frame) splits the steps into two.
}
\end{figure}

To compare this with gapped graphene, the reader is referred to Ref.~\cite{Tabert:2015a}.  The quantization of the Hall conductivity is given by their Eqn.~(15).  To obtain the gapped graphene result, one must take $\Delta_{\rm so}=0$ and $\Delta_z/2\rightarrow\Delta$.  In Fig.~\ref{fig:QHE-G}, the slope of the magnetization is shown for graphene with $\Delta=2$ meV, $E_1/\sqrt{B}=25.64$ meV$/\sqrt{\rm T}$ (a characteristic value for graphene) and $B=1$T.  
The upper two frames show the contribution from the $K$- and $K^\prime$-point, respectively.  The lower frame shows the total result.  Note that the spin-degeneracy is included.  For a single spin species, the Hall response is characterized by a half integer filling factor.  The location of the $N=0$ LL is symmetric between the two valleys (sitting at $-\Delta$ for $K$ and $\Delta$ for $K^\prime$).  As a result, the total system near $\mu=0$ is insulating ($\sigma_H=0$) since the individual valleys contribute equal and opposite edge channels.  Here, a Zeeman interaction (dashed purple curve in the upper frame) splits the steps in the conductivity into two.

\section{Magnetic Oscillations}

We now wish to explore the quantum oscillations which exist for low $B$ fields.  This can be done by extracting the oscillating part of the density of states, using that to find the grand potential and taking the appropriate derivative ($M=-\partial\Omega/\partial B$).  To leading order in $1/(mv_F^2)$, the magnetization is (see Appendix)
\begin{align}\label{Mosc}
M_{\rm osc}^k(\mu)\approx -\frac{e}{2\pi k h}&\frac{(\mu^2-\Delta^2)}{\mu}\left[1+\frac{\mu^2-\Delta^2}{2\mu mv_F^2}\right]{\rm sin}\left(2\pi kx_1\right),
\end{align}
where
\begin{align}\label{x1}
x_1\approx\frac{\displaystyle \left(1-\frac{\mu}{mv_F^2}\right)\left[\mu^2-\Delta^2\right]}{2E_1^2}+\frac{\Delta(1+g)}{2mv_F^2}.
\end{align}

Comparing Eqn.~\eqref{x1} to the customary\cite{Luk:2004}
\begin{align}
x_1=\frac{\hbar A(\mu)}{2\pi eB}-\gamma,
\end{align}
the coefficient of the $E_1^{-2}\propto 1/B$ term in Eqn.~\eqref{x1} is in fact related to the area of the cyclotron orbit:
\begin{align}\label{Area}
A(\mu)=\pi k_F^2\approx\frac{\pi\left[\mu^2-\Delta^2\right]}{\hbar^2v_F^2}\left(1-\frac{\mu}{mv_F^2}\right).
\end{align}
The remainder is the phase shift which is independent of $B$. It is,
\begin{align}\label{gamma}
\gamma=-\frac{\Delta(1+g)}{2mv_F^2}.
\end{align}
In the pure Dirac limit ($m\rightarrow\infty$, $\Delta\rightarrow 0$), $A(\mu)$ reduces to the correct value\cite{Luk:2004,Suprunenko:2008} of $\pi\mu^2/(\hbar^2 v_F^2)$.  For finite $m$ and $\Delta=0$, we obtain the correction of $-[\pi\mu^2/(\hbar^2 v_F^2)][\mu/(mv_F^2)]$ to $A(\mu)$\cite{Tabert:2015}.  Clearly, the gapless limit has a phase shift of 0 associated with a Berry's phase of $\pi$\cite{Koshino:2008, Tabert:2015}.  The amplitude of the quantum oscillations [in units of $-e/(2\pi kh)$] is
\begin{align}
\mathcal{A}_M=\frac{(\mu^2-\Delta^2)}{\mu}\left[1+\frac{\mu^2-\Delta^2}{2\mu mv_F^2}\right]
\end{align}
which properly reduces to the results of Sharapov et al\cite{Sharapov:2004} when $m\rightarrow\infty$, i.e. $\mathcal{A}_M=(\mu^2-\Delta^2)/\mu$ (see their Eqn.~(8.10) in the pure limit), and to the results of Tabert and Carbotte\cite{Tabert:2015} when $\Delta=0$, i.e. $\mathcal{A}_M=\mu[1+\mu/(2mv_F^2)]$ [see their Eqn.~(49)].  Wright and McKenzie\cite{Wright:2013} applied the semiclassical quantization method of Onsager\cite{Onsager:1952}.  This was augmented by including a first correction in the band structure for the energy shift due to the magnetic response of the system (as discussed by Fuchs et al\cite{Fuchs:2010}).  They obtained a phase offset in the magnetic oscillations which, to within a sign that can be traced to the chirality of the Hamiltonian, reduces to our result [Eqn.~\eqref{gamma}] for large $m$ and when the Zeeman interaction is neglected (i.e. $g=0$).
 
\section{Conclusions}

The simplest description of the surface states of a TI is a single Dirac cone centred at the $\Gamma$-point of the 2D surface Brillouin zone.  The presence of a subdominant quadratic-in-momentum Schr{\"o}dinger term to the purely relativistic linear-in-momentum Dirac Hamiltonian is also an important feature; it reshapes the perfect graphene-like cone into an hourglass shape with the narrowing of the conduction cone and outward fanning of the lower cone (valence band). Particle-hole symmetry is lost which can have important ramifications on physical properties such as the density of states, magnetization and the optical properties (both AC and DC).

Doping the surface of a TI with magnetic atoms breaks time-reversal symmetry and creates a gap of $2\Delta$ at the Dirac point.  Alternatively, a slab can be made thin enough that the top and bottom surface states hybridize; consequently, both become gapped.  The sign of the gap is opposite on the two surfaces.  We have studied the magneto-optical response as a function of energy of such systems including as well the effect of a Zeeman interaction.  Particular emphasis is given to the particle-hole asymmetry brought about by the Schr{\"o}dinger mass term associated with the non-relativistic piece of the Hamiltonian.  In a finite magnetic field, both the Zeeman interaction and the gap can modify this asymmetry.  Comparing the integrated density of states $I(\omega_{\rm max})$ to $\omega_{\rm max}$ in a TI to a single valley of graphene highlights the important differences between the two cases.  Namely, particle-hole asymmetry and the very different effect of Zeeman splitting.  For a TI, Zeeman coupling does not split the steps of $I(\omega_{\rm max})$ into two spin-polarized substeps displaced by a constant amount as in graphene.  Instead, it shifts the onset of the various steps by a level-dependent amount.   For a single valley of graphene, the gap does produce asymmetry; but, this will not be seen if the two valleys of opposite chirality are superimposed.  Since the Hamiltonian for a TI involves real spin (as opposed to pseudospin in graphene and related materials such as MoS$_2$ and silicene), only the $N=0$ level is fully spin-polarized\cite{Vazifeh:2012}.  All other LLs are found to have a much reduced value of the average $\hat{s}_z\sim(\hbar/2)[\Delta-(1+g)E_0/2]/(E_1\sqrt{2N})$.  This is to be contrasted with the pseudospin case where all levels are fully $\hat{s}_z$ polarized with $s_z=\pm\hbar/2$ depending on whether the LL has moved up or down in energy due to Zeeman.

The dynamic longitudinal magneto-optical conductivity for a TI also displays asymmetry between positive and negative values of the chemical potential $\mu$.  Even for $\mu=0$, the absorption line associated with the $\mathcal{E}_{1,-}$ to $\mathcal{E}_{0,+}$ transition can be eliminated and replaced by a $\mathcal{E}_{0,+}$ to $\mathcal{E}_{1,+}$ line by manipulating the gap value.  There are peaks in the imaginary part of the transverse Hall AC conductivity which do not exist for the pure Dirac case.  These features do not depend on the subdominant non-relativistic term being large.  They correspond to the interband optical transitions and come as pairs of peaks (one positive the other negative).  They translate into new absorption peaks for circularly polarized light.

While the starting formulas for the magnetization and the DC limit for the real part of the Hall conductivity are quite different, the values of the Hall plateaus that ultimately result are identical and agree with those for the pure relativistic case.  Although, the values of chemical potential at which a new step appears are different.  They depend on the presence of a non-relativistic piece in the Hamiltonian.  For example, the transition form $-1/2$ to $1/2$ no longer corresponds to zero chemical potential.  This transition is shifted to the position of the $N=0$ LL which is given by $(1+g)E_0/2-\Delta$.  We emphasize that this quantity remains finite when both the gap ($\Delta$) and Zeeman term ($g$) are zero.  The shift of $E_0/2$ is directly due to the non-relativistic Schr{\"o}dinger term in our Hamiltonian.  Even if this term is small, it provides a qualitative modification of the physics as compared with the pure relativistic case.  Note that including a Zeeman term and the gap further shifts the energy at which the Hall conductivity transitions from $-1/2$ to $1/2$.  It also depends on the sign of the gap.  For $\Delta>0$, the shift is to lower energies while for $\Delta<0$, it is toward higher energies.

We consider the quantum oscillations that arise in the magnetization.  Unlike previous work, our considerations do not involve any semiclassical arguments based on Onsager's quantization condition for cyclotron orbits.  Here we proceed from the grand potential and use a Poisson formula to obtain the low-field limit $(B\rightarrow 0)$.  We find that when the gap is finite and a Schr{\"o}dinger mass term is also included, there is an offset in the phase shift associated with the magnetic oscillations.  To lowest order, it is given by $\gamma\equiv-\Delta(1+g)/(2mv_F^2)$ which reduces to zero when $\Delta= 0$ or $m\rightarrow\infty$.  In both these cases, $\gamma=0$ as expected for Dirac fermions.  We also see a dependence on the Zeeman splitting $(g)$.  In this formula, $v_F$ is the Dirac Fermi velocity.  Except for a sign due to the chirality of the Hamiltonian, this agrees with Eqn.~(26) in Ref.~\cite{Wright:2013} when only the leading order in $1/m$ is retained.  These authors employed a semiclassical approximation to obtain their results.  We note that this phase offset remains even though the quantization of the Hall plateaus is relativistic ($\pm 1/2, \pm 3/2, \pm 5/2,...$).  A new expression for the amplitude of the magnetic oscillations is given which properly reduces to that of gapped graphene when the subdominant Schr{\"o}dinger term is dropped and to that previously found in Ref.~\cite{Tabert:2015} when the Schr{\"o}dinger mass is included but the gap is set to zero.

\begin{acknowledgments}
We thank E. J. Nicol for support of this work.  This work has been supported by the Natural Sciences and Engineering Research Council of Canada and, in part, by the Canadian Institute for Advanced Research.
\end{acknowledgments}

\appendix*

\section{}

To obtain the magnetic oscillations, return to Eqn.~\eqref{DOS} and express it as
\begin{align}\label{DOS-dw}
N(\omega)=\frac{eB}{h}\frac{d}{d\omega}\left[\Theta\left(\omega-\mathcal{E}_{0,+}\right)+\sum_{\substack{N=1 \\ s=\pm}}^\infty\Theta\left(\omega-\mathcal{E}_{N,s}\right)\right].
\end{align}
The oscillating part of the density of states can be extracted by applying the Poisson formula
\begin{align}
\sum_{N=1}^\infty F(N)=-\frac{1}{2}&F(0)+\int_0^\infty F(x)dx\\
&+2\sum_{k=1}^\infty\int_0^\infty F(x){\rm cos}(2\pi kx)dx.\notag
\end{align}
Following arguments given by Suprunenko et al\cite{Suprunenko:2008}, this gives,
\begin{align}\label{DOS-Osc}
N(\omega)&=\frac{eB}{h}\frac{d}{d\omega}\left\lbrace\frac{1}{2}\Theta(\omega-|\mathcal{E}_{0,+}|)-\frac{1}{2}\Theta(\omega+|\mathcal{E}_{0,+}|)\right.\notag\\
&+\left[\Theta(\omega+|\mathcal{E}_{0,+}|)+\Theta(\omega-|\mathcal{E}_{0,+}|)-\Theta(\omega-\omega_{\rm min})\right]\notag\\
&\times\left[x_1+\sum_{k=1}^\infty\frac{1}{\pi k}{\rm sin}(2\pi kx_1)\right]\notag\\
&+\left. \Theta(\omega-\omega_{\rm min})\left[x_2+\sum_{k=1}^\infty\frac{1}{\pi k}{\rm sin}(2\pi kx_2)\right]\right\rbrace,
\end{align}
where
\begin{align}
\omega_{\rm min}=-\frac{E_1^2}{2E_0}-\frac{E_0}{2E_1^2}\left(\frac{E_0}{2}(1+g)-\Delta\right)^2,
\end{align}
and
\begin{align}\label{xi}
x_i=\frac{E_1^2}{E_0^2}+\frac{\omega}{E_0}+(-)^i\sqrt{\frac{E_1^4}{E_0^4}+\frac{2E_1^2\omega}{E_0^3}+\left(\frac{1+g}{2}-\frac{\Delta}{E_0}\right)^2},
\end{align}
with $i=1,2$.  This generalizes Eqn.~(8) of Ref.~\cite{Suprunenko:2008} to include a gap and Zeeman splitting.  For $E_1^2\gg E_0^2$ and $m\rightarrow\infty$,
\begin{align}
x_1\approx\frac{\omega^2}{2E_1^2}\left[1-\frac{\omega}{mv_F^2}\right]-\frac{bE_0}{2mv_F^2}+\frac{\omega b E_0}{(mv_F^2)^2},
\end{align}
where
\begin{align}
b\equiv\frac{\left[\Delta-(1+g)E_0/2\right]^2}{E_0^2},
\end{align}
and, thus
\begin{align}
-\frac{bE_0}{2mv_F^2}&=-\frac{1}{2mv_F^2E_0}\bigg[\Delta^2-E_0\Delta(1+g)\notag\\
&+\left(\frac{E_0}{2}\right)^2(1+g)^2\bigg].
\end{align}
The last term in the above expression goes like $E_0\propto B$ and will drop out in the limit $B\rightarrow 0$\cite{Tabert:2015}.  This leaves
\begin{align}
-\frac{bE_0}{2mv_F^2}&=-\frac{\Delta^2}{2E_1^2}+\frac{\Delta(1+g)}{2mv_F^2},
\end{align}
where the last term is constant in $B$ and will therefore contribute a constant phase to the quantum oscillations.  We now have
\begin{align}
x_1\approx\frac{1}{2E_1^2}\left[1-\frac{\omega}{mv_F^2}\right]\left[\omega^2-\Delta^2\right]+\frac{\Delta(1+g)}{2mv_F^2}\left[1-\frac{\omega}{mv_F^2}\right].
\end{align}
To lowest order in $1/(mv_F^2)$, 
\begin{align}\label{x1-4app}
x_1\approx\frac{1}{2E_1^2}\left[1-\frac{\omega}{mv_F^2}\right]\left[\omega^2-\Delta^2\right]+\frac{\Delta(1+g)}{2mv_F^2}.
\end{align}
As we do not allow the valence band to bend back toward the zero energy axis, and we focus on $\omega>0$, $x_2$ can be ignored and the oscillating part of the density of states is entirely determined by
\begin{align}\label{DOS-x1}
N_{\rm osc}(\omega)&=\frac{eB}{h}\frac{d}{d\omega}\Theta(\omega-|\mathcal{E}_{0,+}|)\sum_{k=1}^\infty\frac{1}{\pi k}{\rm sin}(2\pi kx_1).
\end{align}
All the information about the quantum oscillations is contained in the first term of Eqn.~\eqref{Omega-minus-vac}.  The oscillating part of the grand thermodynamic potential is thus,
\begin{align}
\Omega_{\rm osc}(\mu)=\int_0^\mu (\omega-\mu)N_{\rm osc}(\omega)d\omega.
\end{align}
where
\begin{align}
N_{\rm osc}(\omega)&=\frac{eB}{h}\sum_{k=1}^\infty\frac{d}{d\omega}N_{\rm osc}^k(\omega),
\end{align}
and
\begin{align}\label{DOS-x1}
N_{\rm osc}^k(\omega)&=\Theta(\omega-|\mathcal{E}_{0,+}|)\frac{{\rm sin}(2\pi kx_1)}{\pi k}.
\end{align}
Therefore,
\begin{align}\label{Omega-k}
\Omega_{\rm osc}^k(\mu)&=-\frac{eB}{\pi k h}\int_0^\mu\Theta(\omega-|E_{0,+}|){\rm sin}(2\pi kx_1)d\omega.
\end{align}
The magnetization is given by $M=-\partial\Omega/\partial B$.  Using
\begin{align}
\frac{\partial x_1}{\partial B}=-\frac{1}{2E_1^2B}\left[1-\frac{\omega}{mv_F^2}\right]\left[\omega^2-\Delta^2\right],
\end{align}
the magnetization can be written as
\begin{align}
M_{\rm osc}^k(\mu)&=\frac{e}{\pi k h}\int_{|\Delta-(1+g)E_0/2|}^\mu \bigg[{\rm sin}(2\pi kx_1)\\
&-\frac{\pi k}{E_1^2}\left[1-\frac{\omega}{mv_F^2}\right]\left[\omega^2-\Delta^2\right]{\rm cos}(2\pi kx_1)\bigg]d\omega.\notag
\end{align}
In the limit of interest ($E_0\rightarrow 0$), the lower bound of integration can be replaced by $\Delta$.  To solve the integral, define
\begin{align}
y&\equiv\left[1-\frac{\omega}{mv_F^2}\right]\left[\omega^2-\Delta^2\right]\notag\\
\implies dy&=\left[2\omega -\frac{3\omega^2}{mv_F^2}+\frac{\Delta^2}{mv_F^2}\right]d\omega\notag\\
\implies d\omega &=\frac{dy}{\displaystyle 2\omega-\frac{3\omega^2}{mv_F^2}+\frac{\Delta^2}{mv_F^2}}.
\end{align}
For $m\rightarrow\infty$, $y\approx\omega^2-\Delta^2$ so, to a first order correction in $1/m$,
\begin{align}
y&\approx \left[1-\frac{\sqrt{y+\Delta^2}}{mv_F^2}\right]\left[\omega^2-\Delta^2\right]\notag\\
\end{align}
Expanding further, we obtain
\begin{align}
\omega &\approx \sqrt{y\left[1+\frac{\sqrt{y+\Delta^2}}{mv_F^2}\right]+\Delta^2},
\end{align}
and
\begin{align}
\omega^2 &\approx y\left[1+\frac{\sqrt{y+\Delta^2}}{mv_F^2}\right]+\Delta^2.
\end{align}
Therefore,
\begin{align}
\frac{1}{\displaystyle 2\omega-\frac{3\omega^2}{mv_F^2}+\frac{\Delta^2}{mv_F^2}}\approx \frac{1}{2\sqrt{y+\Delta^2}}\left[1+\frac{\sqrt{y+\Delta^2}}{mv_F^2}\right].
\end{align}
Consolidating these results, 
\begin{align}
M_{\rm osc}^k(\mu)&\approx\frac{e}{2\pi k h}\int_0^\alpha \left\lbrace{\rm sin}\left(2\pi k\left[\frac{y}{2E_1^2}+\delta\right]\right)\right.\notag\\
&\left.-\frac{\pi k y}{E_1^2}{\rm cos}\left(2\pi k\left[\frac{y}{2E_1^2}+\delta\right]\right)\right\rbrace\notag\\
&\quad\quad\quad\quad\times\left[\frac{1}{\sqrt{y+\Delta^2}}+\frac{1}{mv_F^2}\right]dy.
\end{align}
where
\begin{align}
\alpha\equiv \left[1-\frac{\mu}{mv_F^2}\right]\left[\mu^2-\Delta^2\right],
\end{align}
and
\begin{align}
\delta\equiv\frac{\Delta(1+g)}{2mv_F^2}.
\end{align}
Next, define $y\equiv\alpha x$ to obtain
\begin{align}
M_{\rm osc}^k(\mu)\approx &\frac{e}{2\pi k h}\int_0^1 dx\left\lbrace \frac{\alpha}{\sqrt{\alpha x+\Delta^2}}+\frac{\alpha}{mv_F^2}\right\rbrace\notag\\
&\times\left\lbrace{\rm sin}\left(ax+\bar{\delta}\right)-ax{\rm cos}\left(ax+\bar{\delta}\right)\right\rbrace,
\end{align}
where $a\equiv\pi k\alpha/E_1^2$, $\bar{\delta}\equiv 2\pi k \delta$, and again we work to lowest order in $1/(mv_F^2)$.  To proceed, consider
\begin{align}
I_1&\equiv \int_0^1\left\lbrace{\rm sin}\left(ax+\bar{\delta}\right)-ax{\rm cos}\left(ax+\bar{\delta}\right)\right\rbrace \frac{dx}{\sqrt{x+\Delta^2/\alpha}}\notag\\
&=\int_0^1 {\rm sin}\left(ax+\bar{\delta}\right)\frac{dx}{\sqrt{x+\Delta^2/\alpha}}\notag\\
&\quad\quad-\int_0^1\frac{x}{\sqrt{x+\Delta^2/\alpha}}dx\frac{d}{dx}{\rm sin}\left(ax+\bar{\delta}\right).
\end{align}
Integrating the second term by parts and defining $\omega\equiv x+\Delta^2/\alpha$ and $b\equiv\bar{\delta}-a\Delta^2/\alpha$, we obtain
\begin{align}\label{I1-start}
I_1=\int_{\Delta^2/\alpha}^{1+\Delta^2/\alpha} {\rm sin}\left(a\omega+b\right)&\left[\frac{3}{2\sqrt{\omega}}+\frac{\Delta^2/\alpha}{2\omega^{3/2}}\right]d\omega\notag\\
&-\frac{{\rm sin}\left(a+\bar{\delta}\right)}{\sqrt{1+\Delta^2/\alpha}}.
\end{align}
To solve the remaining integral, we use the definition of the Fresnel sine and cosine integrals:
\begin{align}
\mathcal{S}(z)\equiv\int_0^z{\rm sin}\left(\frac{1}{2}\pi t^2\right)dt,
\end{align}
and
\begin{align}
\mathcal{C}(z)\equiv\int_0^z{\rm cos}\left(\frac{1}{2}\pi t^2\right)dt,
\end{align}
respectively.  We make use of the fact that
\begin{align}
{\rm sin}\left(a\omega+b\right)={\rm sin}(a\omega){\rm cos}b+{\rm cos}(a\omega){\rm sin}b.
\end{align}
We have
\begin{align}\label{1star}
\int_{\Delta^2/\alpha}^{1+\Delta^2/\alpha} {\rm sin}\left(a\omega+b\right)&\frac{d\omega}{\sqrt{\omega}}\notag\\
=\sqrt{\frac{2\pi}{a}}&\left[\mathcal{S}\left(\sqrt{\frac{2a}{\pi}}\omega\right)\Big|^{\sqrt{1+\Delta^2/\alpha}}_{\sqrt{\Delta^2/\alpha}}{\rm cos}b\right.\notag\\
&\left.+\mathcal{C}\left(\sqrt{\frac{2a}{\pi}}\omega\right)\Big|^{\sqrt{1+\Delta^2/\alpha}}_{\sqrt{\Delta^2/\alpha}}{\rm sin}b\right].
\end{align}
Once evaluated at the limits, we will have Fresnel functions with arguments proportional to $\sqrt{a}$.  We are interested in the limit $a\rightarrow\infty$.  We use the asymptotic expansions of the Fresnel integrals for $a\rightarrow\infty$:
\begin{align}
\mathcal{S}\left(\gamma\sqrt{a}\right)\approx\frac{1}{2}-\frac{1}{\pi\gamma\sqrt{a}}{\rm cos}\left(\frac{1}{2}\pi\gamma^2a\right),
\end{align}
and
\begin{align}
\mathcal{C}\left(\gamma\sqrt{a}\right)\approx\frac{1}{2}+\frac{1}{\pi\gamma\sqrt{a}}{\rm sin}\left(\frac{1}{2}\pi\gamma^2a\right).
\end{align}
Therefore, Eqn.~\eqref{1star} is proportional to $1/a$ and is negligible in the limit of interest.  We now return to Eqn.~\eqref{I1-start} and consider
\begin{align}
&\int_{\Delta^2/\alpha}^{1+\Delta^2/\alpha} {\rm sin}\left(a\omega+b\right)\frac{d\omega}{\omega^{3/2}}\notag\\
&=-\frac{2}{\sqrt{\omega}}{\rm sin}(a\omega+b)\Big|^{1+\Delta^2/\alpha}_{\Delta^2/\alpha}\notag\\
&-2\sqrt{2\pi a}\left[\mathcal{S}\left(\sqrt{\frac{2a}{\pi}}\omega\right){\rm sin}b-\mathcal{C}\left(\sqrt{\frac{2a}{\pi}}\omega\right){\rm cos}b\right]\Big|^{\sqrt{1+\Delta^2/\alpha}}_{\sqrt{\Delta^2/\alpha}}.
\end{align}
Applying the Fresnel expansions for $a\rightarrow \infty$, this term is zero.  We arrive at the simple result
\begin{align}\label{I1}
I_1\approx -\frac{{\rm sin}(a+\bar{\delta})}{\sqrt{1+\Delta^2/\alpha}}.
\end{align}

Next, we consider
\begin{align}
I_2&\equiv \int_0^1\left\lbrace{\rm sin}\left(ax+\bar{\delta}\right)-ax{\rm cos}\left(ax+\bar{\delta}\right)\right\rbrace dx\notag\\
&=\int_0^1 {\rm sin}\left(ax+\bar{\delta}\right)dx-\int_0^1 xdx\frac{d}{dx}{\rm sin}\left(ax+\bar{\delta}\right).
\end{align}
Integrating the second term by parts, we obtain
\begin{align}
I_2&=-\frac{2}{a}{\rm cos}\left(ax+\bar{\delta}\right)\Big|_0^1-{\rm sin}\left(a+\bar{\delta}\right).
\end{align}
In the limit $a\rightarrow\infty$,
\begin{align}\label{I2}
I_2\approx -{\rm sin}(a+\bar{\delta}).
\end{align}
Combing Eqns.~\eqref{I1} and \eqref{I2}, the magnetization becomes
\begin{align}
M_{\rm osc}^k(\mu)\approx -\frac{e}{2\pi k h} &\left[\frac{\alpha}{\sqrt{\alpha+\Delta^2}}+\frac{\alpha}{mv_F^2}\right]{\rm sin}(a+\bar{\delta}).
\end{align}
Written in terms of the original variables,
\begin{align}\label{Mosc-app}
M_{\rm osc}^k(\mu)\approx -\frac{e}{2\pi k h}&\frac{(\mu^2-\Delta^2)}{\mu}\left[1+\frac{\mu^2-\Delta^2}{2\mu mv_F^2}\right]{\rm sin}\left(2\pi kx_1\right),
\end{align}
where
\begin{align}\label{x1-app}
x_1\approx\frac{\displaystyle \left(1-\frac{\mu}{mv_F^2}\right)\left[\mu^2-\Delta^2\right]}{2E_1^2}+\frac{\Delta(1+g)}{2mv_F^2}.
\end{align}

\bibliographystyle{apsrev4-1}
\bibliography{TI-GAP}

\end{document}